\documentclass[%
 reprint,
 amsmath,amssymb,
 aps,
]{revtex4-1}

\usepackage{graphicx}% Include figure files
\usepackage{dcolumn}% Align table columns on decimal point
\usepackage{bm}% bold math
\usepackage[colorlinks,linkcolor=blue,anchorcolor=blue,citecolor=blue,urlcolor=black]%
{hyperref}
\begin{document}

\preprint{APS/123-QED}

\title{Simulating the Lipkin-Meshkov-Glick model in a hybrid quantum system }% Force line breaks with with nitrogen-vacancy centers in diamond coupled to microwave superconducting cavity

\author{Yuan Zhou}
\author{Sheng-Li Ma}
\author{Bo Li}
\author{Xiao-Xiao Li}
\author{Fu-Li Li}
\author{Peng-Bo Li}%
 \email{lipengbo@mail.xjtu.edu.cn}
\affiliation{Shaanxi Province Key Laboratory of Quantum Information and Quantum Optoelectronic Devices,
Department of Applied Physics, Xi¡¯an Jiaotong University, Xi¡¯an 710049, China
}%

\date{\today}% It is always \today, today,

\begin{abstract}
We propose an efficient scheme for simulating the Lipkin-Meshkov-Glick (LMG) model with nitrogen-vacancy (NV) center ensembles in diamond magnetically coupled to superconducting coplanar waveguide cavities. With the assistance of external microwave driving fields, we show that the interaction of the NV spins can be easily controlled, and several types of the LMG model can be realized by tuning the different parameters. Under the thermal dynamical limit, the distinct non-equilibrium second order quantum phase transition of the spin ensemble can be achieved at the critical point. Furthermore, we show that the spin squeezed state can be generated by tailoring the LMG Hamiltonian to possess the two-axis counter-twisting form in this hybrid quantum system.

%\begin{description}
%42.50.Dv, 03.67.-a, 76.30.Mi
%\end{description}
\end{abstract}

\maketitle

%\tableofcontents

\section{introduction}

Manipulating the couplings of collective particles has been a
fascinating subject with the development of new technologies for
ultracold atoms, trapped ions, and solid-state spins \cite{1,2,3,4,5,6,7,8,9,10,11}.
One of the most important applications is to simulate the Lipkin-Meshkov-Glick (LMG) model in these different systems \cite{12,13,14}.
This model, which was first proposed in nuclear physics \cite{15}, has become a hot issue in the field of quantum information and quantum simulation \cite{16,17,18}.
Because of utilizing this model, we can not only manipulate the special quantum states such as the coherent spin state or spin squeezed state \cite{19,20,21,22}, but also \textquotedblleft tailor\textquotedblright\ the microscopic interaction between the particles to mimic quantum phase transitions of the macroscopic system \cite{16,17,18,23}.
In spite of many outstanding investigations for simulating the LMG model, it remains a challenge to realize the general LMG model in laboratory \cite{20,21,22,23}.
Therefore, it is appealing to present an experimentally feasible scheme for realizing the LMG model.

Recently, much attention has been paid to manipulating nitrogen-vacancy (NV) center ensembles in hybrid quantum systems \cite{24,25,26,27,28,29,30,31,32,33,34,35,36,37,38,39}.
NV centers in diamond have exhibited the excellent features such as fast microwave manipulation, optical
preparation, and detection, and long coherence time even
at room temperature \cite{40,41,42,43,44,45,46,47,48}.
Besides, we can directly combine NV centers with other quantum systems without requiring sophisticated trapping techniques \cite{40,41}. Therefore, hybrid quantum systems composed of superconducting circuits and NV centers have been extensively investigated \cite{41,42,43,44,45,46,47,48,49,50,51,52,53,54,55,56}.
Utilizing these hybrid systems, one can prepare fantastic quantum states, design quantum logic gates, store or transfer quantum states \cite{24,25,26,27,28,29,31,32,33,34,35,44,45,46,47,48,49,51,56}.
Furthermore, we can perform some quantum simulating tasks with this spin-photon system \cite{33,57,58}. In a recent paper,
a protocol for simulating the Dicke model and Dicke Lattice Model is proposed with the isotropy and anisotropy NV center ensembles in the periodic superconducting  microwave cavities respectively \cite{59}.
This prominent work demonstrates that hybrid quantum
systems provide a realistic platform for studying characteristic phenomena of nonequilibrium quantum systems in various configurations.

In this work, we propose an experimentally feasible scheme for
simulating the LMG model in a hybrid quantum system with an NV center ensemble in diamond coupled to superconducting coplanar waveguide cavities. Under the condition of large detunings as well as the bad cavity limit, we can obtain the generalized LMG model in this hybrid system. We discuss several forms of the LMG model by adjusting the parameters such as detunings, Rabi frequencies, coupling coefficients and so on. In particular, we focus on the positive field case $h>0$ of the $\chi=0$ LMG model with ferromagnetic interactions $\lambda >0$.  Under the thermal dynamical limit, the distinct non-equilibrium second order quantum phase transition of the coupled spins can be achieved at the critical point, as the magnitude of the interaction strength varies. On the other hand, the LMG model with the form of $H\sim \hat{J}_{x}^{2},\hat{J}_{y}^{2},\hat{J}_{z}^{2} $ or $H\sim (\hat{J}_{x}^{2}-\hat{J}_{y}^{2})$, corresponding to the one-axis twisting or two-axis counter-twisting Hamiltonian,
can also be utilized for generating the spin squeezed state \cite{19,20,21,22,23,31,51,58,61}.
Therefore, by tailoring the LMG model with the two-axis counter-twisting form, we can prepare the spin squeezed state with high degree of squeezing based on this kind of interaction. Our work provides a realistic platform for implementing LMG-type models and for studying characteristic phenomena of nonequilibrium quantum systems with hybrid quantum systems.

\section{The model}

\begin{figure}
\includegraphics[width=7cm]{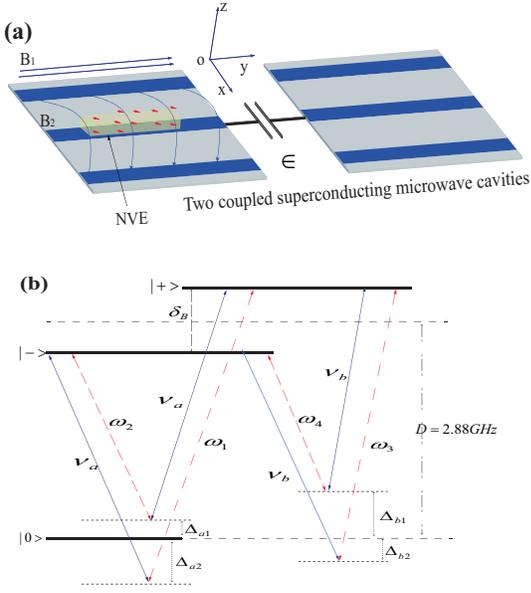}
\caption{\label{fig:wide}(Color online) The scheme diagrams.
(a) Two superconducting coplanar waveguide cavities are
coupled together ( the coupling coefficient is $\epsilon$ ),
an NV center ensemble is set in the center of one cavity and this ensemble
is magnetically coupled to this cavity.
(b) Level diagram of the NV center ground triplet state and the feasible four Raman
 transition channels. The blue solid arrows indicate the
coupling to the two normal modes of the quantized cavities fields with the
frequencies $\nu _{a}$, and $\nu _{b}$. The red dashed arrows show
the four additional classical microwave fields with frequencies $
\omega _{1}$, $\omega_{2}$, $\omega_{3}$, and $\omega_{4}$,
and they are used to implement Raman transitions between the two
excited spin states $|+\rangle $ and $|-\rangle $.}
\end{figure}

As illustrated in Fig.~1(a), in this hybrid quantum system two superconducting coplanar waveguide cavities are strongly coupled together with the coefficient
$\epsilon\sim10$ MHz. Meanwhile one of them is magnetically coupled to
an NV center ensemble \cite{40,41,55,56,57,58,59}. The energy level structure
of the single NV center is shown in Fig.~1(b). The electronic
ground state is the spin triplet state denoted as $
|m_{s}=0,\pm 1\rangle$, and the zero-field splitting between the degenerate
sublevels $|m_{s}=\pm 1\rangle$ and $|m_{s}=0\rangle $ is $D=2\pi \times 2.88$ GHz.
Then we apply a homogeneous static magnetic field $B_{1}$ to remove the degenerate
states $|m_{s}=\pm 1\rangle$ with the Zeeman splitting $\delta _{B}/2\pi=2g_{e}\mu
_{B}B_{1}\sim 100$ MHz $(g_{e}\simeq 2$ is the nitrogen-vacancy land\'{e}
factor, $\mu _{B}=14$ MHz $\text{mT}^{-1}$ is the Bohr magneton$)$,
which results in a three-level system denoted as $|m_{s}=0\rangle \equiv|0\rangle
$, $|m_{s}=+1\rangle \equiv|+\rangle $, and $|m_{s}=-1\rangle \equiv|-\rangle $.

In this solid-state system,
the presence of local strain and the coupling of the NV spins to other electronic or nuclear spins in the
surrounding can substantially generate the random frequency splitting $\delta_{j}$ between the states $|+\rangle$ and $|-\rangle$ for each NV center \cite{59}.
As a result, the coherence time is limited by this inhomogeneous broadening, resulting in an ensemble dephasing time $T^*_{2}$,
which is about one microsecond according to the current experiments \cite{62,63,64}.
However, one can extend this coherence time from $T^*_{2}$ to the value $T_{2}$ (close to the intrinsic spin coherence time)
through the spin echo technology \cite{62,63,64}. It has been reported that, the  NV spin coherence time in an ensemble is comparable to that of single NV center, with $T_{2}>600\mu s$ for a sample with natural abundance of $^{13}C$ and paramagnetic impurity density $\varrho\sim 10^{15} \text{cm}^{-3}$ \cite{65,66}.
In this work, we have neglected  these terms in the Hamiltonian and included their major influence via the dephasing rate $\gamma_\text{dep}$ to the master equation equivalently.

Therefore, the Hamiltonian for describing these two cavities and the homogeneous NV center ensemble is $(\hbar =1)$
\begin{equation}
H_{1}=H_{NVE}+H_{10}+H_{NC}+H_{CC},
\end{equation}%
where $H_{NVE}=\sum_{j=1}^{N}\omega _{+}|+\rangle _{j}\langle +|+\omega_{-}|-\rangle _{j}\langle -|$
is  the free Hamiltonian for the homogeneous NV ensemble,
$H_{10}=\omega _{c1}\hat{c_{1}}^{\dagger }\hat{c_{1}}+\omega _{c2}\hat{c_{2}}^{\dagger
}\hat{c_{2}}$ is the free Hamiltonian for these two cavities with the destruction operators $\hat{c_{i}},(i=1,2)$,
$H_{NC}=\sum_{j=1}^{N}[\eta _{1}\hat{c_{1}}|+\rangle _{j}\langle 0|+\eta
_{2}\hat{c_{1}}|-\rangle _{j}\langle 0|+h.c.]$ is the interaction between
cavity 1 and the NV centers, and $H_{CC}=\epsilon(\hat{c_{1}}\hat{c_{2}}^{\dagger }+\hat{c_{1}}^{\dagger
}\hat{c_{2}})$ is the interaction between these two cavities.
For the related parameters, $N$ is the number of the available NV centers, $\omega _{\pm }=D\pm \delta _{B}/2$ the transition
frequencies between the states $|\pm\rangle $ and $|0\rangle$, $\omega _{c1}$ and $\omega _{c2}$ the cavity mode frequencies, and
$\eta _{1}$ and $\eta _{2}$ are the coupling strengths between the cavity mode $\hat{c_{1}}$ and the j\emph{th} NV center for the transitions $|0\rangle _{j}\rightarrow|+\rangle_{j}$ and $|0\rangle _{j}\rightarrow|-\rangle_{j}$.

In order to acquire four distinct Raman transitions between the two spin states $|m_{s}=+1\rangle $ and $|m_{s}=-1\rangle$, we utilize the canonical transformation $\hat{c_{1}}=(\hat{a}+\hat{b})/\sqrt{2},\hat{c_{2}}=(\hat{a}-\hat{b})/\sqrt{2},\hat{c_{1}}^{\dagger }=(\hat{a}^{\dagger
}+\hat{b}^{\dagger })/\sqrt{2}$, and $\hat{c_{2}}^{\dagger }=(\hat{a}^{\dagger }-\hat{b}^{\dagger })/\sqrt{2}$. As a result, we can get the equivalent form of the above Hamiltonian
\begin{eqnarray}
H_{1}^{^{\prime }} &=&\nu _{a}\hat{a}^{\dagger }\hat{a}+\nu _{b}\hat{b}^{\dagger }\hat{b}  \notag +H_{NVE}\\
&&+\sum_{j=1}^{N}(g_{1}\hat{a}|+\rangle _{j}\langle 0|+g_{2}\hat{a}|-\rangle _{j}\langle
0|  \notag \\
&&+g_{3}\hat{b}|+\rangle _{j}\langle 0|+g_{4}\hat{b}|-\rangle _{j}\langle 0|+\text{H.c.}),
\end{eqnarray}%
where $\hat{a}(\hat{a}^{\dagger })$ and $\hat{b}(\hat{b}^{\dagger })$ are destruction (creation) operators for these two supermodes,
$g_{k}(k=1,2,3,4)$ are the relative average coupling coefficients between the
supermodes and NV centers $(g_{1}=g_{3}=\eta_{1}/\sqrt{2}, g_{2}=g_{4}=\eta_{2}/\sqrt{2})$, and  $\nu _{a}=
\epsilon+(\omega _{c1}+\omega _{c2})/2$ and $\nu _{b}=-\epsilon+(\omega _{c1}+\omega _{c2})/2$
are the relative frequencies of these two supermodes respectively.
Meanwhile, we introduce four microwave classical fields with
frequencies $\omega _{k}(k=1,2,3,4)$ to the NV center ensemble, and neglect the NV centers' anisotropy and the deference from locations.

As a result, the total Hamiltonian for this hybrid quantum system can be expressed as
\begin{equation}
H_\text{total}=H_{0}+H_\text{int},
\end{equation}%
where
\begin{equation}
H_{0}=\nu _{a}\hat{a}^{\dagger }\hat{a}+\nu _{b}\hat{b}^{\dagger
}\hat{b}+\sum_{j=1}^{N}(\omega _{+}|+\rangle _{j}\langle +|+\omega _{-}|-\rangle
_{j}\langle -|)\notag,
\end{equation}
\begin{eqnarray}
H_\text{int} &=&\sum_{j=1}^{N}(g_{1}\hat{a}|+\rangle _{j}\langle 0|+g_{2}\hat{a}|-\rangle_{j}\langle 0| \notag \\
&&+g_{3}\hat{b}|+\rangle _{j}\langle 0|+g_{4}\hat{b}|-\rangle _{j}\langle 0|  \notag \\
&&+\Omega _{1}/2|+\rangle
_{j}\langle 0|e^{-i\omega _{1}t}+\Omega _{2}/2|-\rangle _{j}\langle
0|e^{-i\omega _{2}t}  \notag \\
&&+\Omega _{3}/2|+\rangle _{j}\langle 0|e^{-i\omega _{3}t}+
\Omega _{4}/2|-\rangle _{j}\langle 0|e^{-i\omega _{4}t}+\text{H.c.})\notag .
\end{eqnarray}%
In the interaction picture, we have
\begin{eqnarray}
H_{I} &=&\sum_{j=1}^{N}(g_{1}\hat{a}|+\rangle _{j}\langle 0|e^{i\Delta _{a1}t}+g_{2}\hat{a}|-\rangle
_{j}\langle 0|e^{-i\Delta _{a2}t}\notag\\&&
+g_{3}\hat{b}|+\rangle _{j}\langle 0|e^{i\Delta _{b1}t}+g_{4}\hat{b}|-\rangle _{j}\langle 0|e^{-i\Delta _{b2}t}\\&&
+\Omega _{1}/2|+\rangle_{j}\langle 0|e^{-i\Delta _{a2}t}+\Omega _{2}/2|-\rangle _{j}\langle 0|e^{i\Delta _{a1}t} \notag \\&&
+\Omega _{3}/2|+\rangle _{j}\langle 0|e^{-i\Delta _{b2}t}+\Omega _{4}/2|-\rangle _{j}\langle 0|e^{i\Delta _{b1}t}+\text{H.c.})\notag ,
\end{eqnarray}%
where $(\Delta _{a1}+\Delta _{a2})\sim(\Delta _{b1}+\Delta _{b2})\sim\delta_{B}$, $\Delta _{a1}=\omega _{+}-\nu_{a}=\omega _{-}-\omega _{2}$, $\Delta _{a2}=\omega _{1}-\omega _{+}=\nu_{a}-\omega _{-}$, $\Delta _{b1}=\omega _{+}-\nu_{b}=\omega _{-}-\omega _{4}$, and $\Delta _{b2}=\omega _{3}-\omega _{+}=\nu_{b}-\omega _{-}$.
We assume that all of the fields are on the two-photon resonance in these four Raman transitions and take advantage of the relations $\delta_{B}\gg g_{i},\Omega_{i}$, $|\Delta _{a1}\pm \Delta _{b1}|\gg g_{i},\Omega_{i}$, and $|\Delta _{a2}\pm \Delta _{b2}|\gg g_{i},\Omega_{i}$. In this case, the total effective Hamiltonian can be written as \cite{60}
\begin{eqnarray}
H_\text{eff} &=&\mu _{0}\hat{J}_{z}+\zeta _{a}\hat{a}^{\dag }\hat{a}+\zeta _{b}\hat{b}^{\dag
}\hat{b}+\eta _{a}^{-}\hat{J}_{z}\hat{a}^{\dag }\hat{a}+\eta _{b}^{-}\hat{J}_{z}\hat{b}^{\dag }\hat{b}  \notag \\
&&+\dfrac{\sigma _{a}}{\sqrt{N}}(\hat{T}_{a}\hat{a}+\hat{T}_{a}^{\dag }\hat{a}^{\dag })+\dfrac{%
\sigma _{b}}{\sqrt{N}}(\hat{T}_{b}\hat{b}+\hat{T}_{b}^{\dag }\hat{b}^{\dag }),
\end{eqnarray}%
where the effective microscopic parameters are given by
\begin{eqnarray}
&&\mu _{0} =\tfrac{|g_{1}|^{2}}{\Delta _{a1}}-\tfrac{|g_{2}|^{2}}{\Delta _{a2}}+\tfrac{|g_{3}|^{2}}{\Delta _{b1}}-\tfrac{|g_{4}|^{2}}{\Delta _{b2}}\notag\\
&&+\tfrac{1}{4}(\tfrac{|\Omega _{1}|^{2}}{\Delta _{a2}}+\tfrac{|\Omega _{2}|^{2}}{\Delta _{a1}}
-\tfrac{|\Omega _{3}|^{2}}{\Delta _{b2}}-\tfrac{|\Omega _{4}|^{2}}{\Delta _{b1}}),\notag\\
&&\sigma _{a}\alpha _{a}=\tfrac{\sqrt{N}g_{1}\Omega _{2}^{\ast }}{2\Delta _{a1}},\sigma _{a}\beta _{a}=-\tfrac{\sqrt{N}g_{2}\Omega _{1}^{\ast }}{2\Delta _{a2}},\notag\\
&&\sigma _{b}\alpha _{b}=\tfrac{\sqrt{N}g_{3}\Omega _{4}^{\ast }}{2\Delta _{b1}},\sigma _{b}\beta _{b}=-\tfrac{\sqrt{N}g_{4}\Omega _{3}^{\ast }}{2\Delta _{b2}},\notag\\
&&\zeta _{a}=\tfrac{N}{2}(\tfrac{|g_{1}|^{2}}{\Delta _{a1}}-\tfrac{|g_{2}|^{2}}{\Delta _{a2}}),\zeta _{b}=\tfrac{N}{2}(\tfrac{|g_{3}|^{2}}{\Delta _{b1}}-\tfrac{|g_{4}|^{2}}{\Delta_{b2}}),\notag\\
&&\eta _{a}^{-}=(\tfrac{|g_{1}|^{2}}{\Delta _{a1}}-\tfrac{|g_{2}|^{2}}{\Delta _{a2}}),
\eta _{b}^{-}=(\tfrac{|g_{3}|^{2}}{\Delta _{b1}}-\tfrac{|g_{4}|^{2}}{\Delta _{b2}}).
\end{eqnarray}
The operators $\hat{T}_{i}=\alpha _{i}\hat{J}_{+}+\beta _{i}\hat{J}_{-}$, $i=\{a,b\}$, and the dimensionless factors
$\{\alpha _{i},\beta _{i}\}\in [-1,1]$ are introduced for convenience. The coefficients for the free Hamiltonian are $\mu_{0}$, $\zeta_{a}$ and $\zeta_{b}$, and the coefficients for the interactions are $\sigma _{i}\alpha _{i}$ and $\sigma _{i}\beta _{i}(i=\{a,b\})$, the coefficients for the nonlinear items are $\eta_{a}$ and $\eta_{b}$, all of the coefficients above can be tuned by the number of NV centers $N$, the average coupling coefficients $g_{k}$, Rabi frequencies $\Omega^{*}_{k}(k=1,2,3,4)$ and the detunings. The collective ladder operators for the NV center ensemble are
$\hat{J}_{z}=\frac{1}{2}\sum_{j=1}^{N}(|+\rangle _{j}\langle +|-|-\rangle_{j}\langle -|)$,
$\hat{J}_{+}=\sum_{j=1}^{N}|+\rangle _{j}\langle -|$, $\hat{J}_{-}=\sum_{j=1}^{N}|-\rangle _{j}\langle +|$,
and they also satisfy the angular momentum commutation relations
\begin{equation}
\lbrack \hat{J}_{i},\hat{J}_{j}]=i\varepsilon
_{ijk}\hat{J}_{k},[\hat{J}_{+},\hat{J}_{-}]=2\hat{J}_{z},[\hat{J}_{z},\hat{J}_{\pm }]=\mp \hat{J}_{\pm }.
\end{equation}

The master equation for the cavity modes and NV spins is,
\begin{equation}\label{ME1}
\dot{\rho}_{g}=-i[H_\text{eff},\rho _{g}]+\kappa _{a}D[\hat{a}]\rho _{g}+\kappa
_{b}D[\hat{b}]\rho _{g}+\gamma_\text{dep}D[\hat{J_{z}}]\rho _{g},
\end{equation}%
where $D[\hat{O}]\rho =2\hat{O}\rho \hat{O}^{\dag }-\hat{O}^{\dag }\hat{O}\rho -\rho \hat{O}^{\dag }\hat{O}$, $\kappa _{i}$ is the cavity field decay rate,
and $\gamma_\text{dep}$ is the dephasing rate caused by the inhomogeneous broadening.
Now we suppose that the two cavities are both bad cavities and satisfy the relations $\sqrt{\kappa _{i}^{2}+\zeta _{i}^{2}}\gg \sigma
_{i},\mu _{0}$. Therefore, these two supermodes are only   weakly excited and can be adiabatically eliminated from the dynamics \cite{21,23}.
First of all, we set the subspace
for the cavity supermodes as $|0_{a}0_{b}\rangle=|1\rangle$, $|0_{a}1_{b}\rangle=|2\rangle$,
$|1_{a}0_{b}\rangle=|3\rangle$, and $|1_{a}1_{b}\rangle=|4\rangle$, and neglect populations of the highly
excited states. As a result, the master equation (\ref{ME1}) can be expressed as
a set of coupled differential equations for the reduced density matrix elements, with
$\rho =\text{Tr}_\text{fields}(\rho_{g})=(\rho_{11}+\rho_{22}+\rho_{33}+\rho_{44})$. Owing to the strong damping,
the most populated states of the supermodes are in the ground state $|1\rangle$ and the
off-diagonal elements of the reduced density operator change slowly in time, i.e., $\dot\rho_{ij}=0, (i\neq j)$. Taking advantage of the
steady solutions of the off-diagonal elements of $\rho_{ij}$, we have the following master equation for the reduced density operator
\begin{equation}
\dot{\rho}=-i[H,\rho ]+\frac{\Gamma _{a}}{N}D[\hat{T}_{a}^{\dag }]\rho +\frac{%
\Gamma _{b}}{N}D[\hat{T}_{b}^{\dag }]\rho+\gamma_{dep}D[\hat{J_{z}}]\rho ,
\end{equation}%
where
\begin{equation}\label{ME2}
H=\mu _{0}\hat{J}_{z}- \frac{\Lambda _{a}}{N}\hat{T}_{a}\hat{T}_{a}^{\dag }-\frac{\Lambda
_{b}}{N}\hat{T}_{b}\hat{T}_{b}^{\dag } ,
\end{equation}%
with $\Lambda _{i}=\dfrac{\sigma _{i}^{2}\zeta _{i}}{\kappa
_{i}^{2}+\zeta _{i}^{2}}$, and $\Gamma _{i}=\dfrac{\sigma _{i}^{2}\kappa _{i}}{%
\kappa _{i}^{2}+\zeta _{i}^{2}}$. Because of the large cavity dissipations $\kappa_{i}\gg\eta_{a}^{-},\eta_{b}^{-}$,
the nonlinear terms of these two supermodes have no contributions in the adiabatic elimination course.
According to the definition of $\hat{T}_{i}$ and $\hat{T}_{i}^\dag$, we transform the equation (\ref{ME2})
into the generalized LMG model Hamiltonian \cite{15,16,17,18,19,20,21,22,23,61},
\begin{equation}
H_{LMG}=-2h\hat{J}_{z}-\frac{2\lambda }{N}(\hat{J}_{x}^{2}+\chi \hat{J}_{y}^{2}),
\end{equation}%
where the parameters are given by
\begin{equation}
-2h=\mu_{0}-\dfrac{\zeta_{a}((L_{\alpha
}^{a})^{2}-(L_{\beta }^{a})^{2})}{NK_{a}}-\dfrac{\zeta_{b}((L_{\alpha }^{b})^{2}-(L_{\beta }^{b})^{2})}{NK_{b}},\notag
\end{equation}
\begin{equation}
2\lambda=\dfrac{\zeta_{a}(L_{\alpha
}^{a}+L_{\beta }^{a})^{2}}{K_{a}}+\dfrac{\zeta_{b}(L_{\alpha }^{b}+L_{\beta }^{b})^{2}}{K_{b}},\notag
\end{equation}
\begin{equation}
\chi=\dfrac{K_{b}\zeta_{a}(L_{\alpha}^{a}-L_{\beta }^{a})^{2}+K_{a}\zeta_{b}(L_{\alpha}^{b}-L_{\beta }^{b})^{2}}
{K_{b}\zeta_{a}(L_{\alpha}^{a}+L_{\beta }^{a})^{2}+K_{a}\zeta_{b}(L_{\alpha}^{b}+L_{\beta }^{b})^{2}},\notag
\end{equation}
\begin{equation}
K_{a}=(\kappa _{a}^{2}+\zeta_{a}^{2}), K_{b}=(\kappa _{b}^{2}+\zeta_{b}^{2}), \notag
\end{equation}
\begin{equation}
L_{\alpha }^{a}=\sigma _{a}\alpha _{a}, L_{\beta }^{a}=\sigma _{a}\beta_{a}, L_{\alpha }^{b}=\sigma _{b}\alpha _{b}, L_{\beta }^{b}=\sigma _{b}\beta_{b}.\notag
\end{equation}
These parameters can be controlled by adjusting the relevant parameters such as the detunings, Rabi frequencies,
and coupling coefficients.

\begin{table*}
\caption{\label{tab:table1}The relevant  parameters  for the  specific LMG Hamiltonian}
\begin{ruledtabular}
\begin{tabular}{lp{4.6cm}p{3.1cm}p{3.5cm}}
Parameters (MHz)& Two-axis counter-twisting LMG model (Equation (\ref{ME3})) & Isotropic LMG model (Equation (\ref{ME16})) & One-axis twisting LMG model (Equation (\ref{ME4})) \\\hline
$\sqrt{N}g_{k}(N\simeq 10^{12})$ & 12 & 12 & 12 \\
$|\Omega_{1}^*|/2\pi$ & 4 & 0 & 7 \\
$|\Omega_{2}^*|/2\pi$ & 1 & 1 & 3 \\
$|\Omega_{3}^*|/2\pi$ & 1 & 1 & 0.77 \\
$|\Omega_{4}^*|/2\pi$ & 4 & 0 & 0 \\
$|\Delta_{a1}|/2\pi$ & 20 & 20 & 30 \\
$|\Delta_{a2}|/2\pi$ & 80 & 80 & 70 \\
$|\Delta_{b1}|/2\pi$ & 80 & 80 & 50 \\
$|\Delta_{b2}|/2\pi$ & 20 & 20 & 50 \\
$\sigma _{a}\alpha _{a}/2\pi$ & 0.3 & 0.3 & 0.6  \\
$\sigma _{a}\beta _{a}/2\pi$ & 0.3 & 0 & 0.6  \\
$\sigma _{b}\alpha _{b}/2\pi$ & 0.3 & 0 & 0 \\
$\sigma _{b}\beta _{b}/2\pi$ & -0.3 & 0.3 & 0.092 \\%
\end{tabular}
\end{ruledtabular}
\end{table*}

In addition, we consider several forms of the LMG model by tuning these parameters.
In order to describe them more clearly,  we list three groups of parameters in TABLE. \uppercase\expandafter{\romannumeral1}.
Then we can get several different forms of the LMG Hamiltonian via choosing the suitable parameters.
When we set the parameters as those in the first column of TABLE. \uppercase\expandafter{\romannumeral1},
we can get the conventional LMG Hamiltonian with the expression
\begin{equation}\label{ME3}
H=-2h\hat{J}_{z}-\frac{2\lambda }{N}(\hat{J}_{x}^{2}-\hat{J}_{y}^{2}),
\end{equation}%
where $h=-\mu _{0}/2$, $\chi =-1$, $\alpha _{a}=\alpha _{b}=\alpha=\sqrt{2}/2$, $\beta _{a}=-\beta
_{b}=\beta=\sqrt{2}/2$, $\lambda =\Lambda _{a}=-\Lambda _{b}$, and $\sigma _{a}\simeq\sigma _{b}$.
The master equation reduces to the form
\begin{equation}\label{ME15}
\dot{\rho}=-i[H,\rho ]+\frac{2\Gamma _{a}\alpha^{2}}{N}D[\hat{J}_{-}]\rho +\frac{2\Gamma _{b}\beta^{2}}{%
N}D[\hat{J}_{+}]\rho+\gamma_\text{dep}D[\hat{J_{z}}]\rho.
\end{equation}%
This kind of LMG model has been  investigated  for phase transitions and multiparticle entanglement \cite{12,13,14,15,16,17}.
One can generate the spin squeezed state by the two-axis counter-twisting interactions utilizing this kind of Hamiltonian\cite{19,20,21}.

Secondly, we can get the isotropic Hamiltonian when choosing the second column parameters in TABLE. \uppercase\expandafter{\romannumeral1}.
\begin{equation}\label{ME16}
H=-2h\hat{J}_{z}-\frac{2\lambda }{N}(\hat{J}_{x}^{2}+\hat{J}_{y}^{2}),
\end{equation}%
where $h=-\mu _{0}/2$, $\chi =1$, $\Lambda _{a}=\Lambda _{b}\equiv \lambda $, $\alpha _{a}=\beta _{b}=1$, $\alpha_{b}=\beta _{a}=0$ and $\sigma _{a}\simeq\sigma _{b}$.
Then the master equation is
\begin{equation}
\dot{\rho}=-i[H,\rho ]+\frac{\Gamma _{a}}{N}D[\hat{J}_{-}]\rho +\frac{\Gamma _{b}}{%
N}D[\hat{J}_{+}]\rho+\gamma_{dep}D[\hat{J_{z}}]\rho.
\end{equation}
This is an isotropic LMG Hamiltonian which can be solved exactly because of $H=-2h\hat{J}_{z}-2\lambda\hat{\textbf{J}}^2/N+2\lambda\hat{J}_{z}^2/N$, where $\hat{\textbf{J}}^2=\hat{J}_{x}^{2}+\hat{J}_{y}^{2}+\hat{J}_{z}^{2}$.
We can get the double degenerate ground states for this Hamiltonian when $h = 0$.
Otherwise, for the symmetry breaking case $h \neq 0$, we will get an unique ground state
($h > 0$, $|\uparrow\uparrow\uparrow\cdots\rangle=|m_{z}=N/2\rangle$ or $h < 0$, $|\downarrow\downarrow\downarrow\cdots\rangle=|m_{z}=-N/2\rangle$ ).
In addition there will be the transition between the ferromagnetic and antiferromagnetic interactions by tuning the sign of $\lambda$ when we set $h = 0$,
and so on.

Finally, we can obtain the simple Hamiltonian according to the third column in TABLE. \uppercase\expandafter{\romannumeral1},
\begin{equation}\label{ME4}
H=-2h\hat{J}_{z}-\frac{2\lambda }{N}\hat{J}_{x}^{2},
\end{equation}%
and achieve the master equation
\begin{equation}\label{ME5}
\dot{\rho}=-i[H,\rho ]+\frac{\Gamma _{a}}{N}D[2\hat{J}_{x}]\rho +\frac{\Gamma _{b}%
}{N}D[\hat{J}_{+}]\rho+\gamma_\text{dep}D[\hat{J_{z}}]\rho,
\end{equation}%
where $\lambda =2\Lambda _{a}$, $\chi =0$, $\alpha _{a}=\beta _{a}=1$, $\beta_{b}=\sqrt{2}/2$, and $\alpha_{b}=0$.
There will be many other applications of this kind of interactions, e.g., simulating the first order and second order phase transitions,
preparing multiparticle entanglement  and generating spin squeezed state by the one-axis twisting interactions \cite{12,13,14}.
In the following, we will investigate this kind of phase transition in this hybrid quantum system for the Hamiltonian with the form given by equation (\ref{ME4}).

\section{The second-order phase transition}
Under the thermodynamic limit corresponding
to that $N\rightarrow\infty, V\rightarrow\infty$, and the density $n=N/V$ keeps finite,
we can neglect the quantum fluctuation for this spin-spin interaction system,
for example, $\langle \hat{J}_{k}\hat{J}_{l}\rangle \rightarrow
\langle \hat{J}_{k}\rangle \langle \hat{J}_{l}\rangle $, where $k,l\in \{x,y,z\}$.
Considering the ferromagnetic interactions $(\lambda >0)$ from equation (\ref{ME4}),
we will apply the method of
semiclassical equations of motion to simulate the second-order quantum phase transition.
The differential equations of motion for the expectation values of collective spin components
$\langle \hat{J}_{i}\rangle =\text{Tr}(\rho \hat{J}_{i})$ are
readily derived from the master equation (\ref{ME5}).
Utilizing the relations $d\langle\hat{J}_{i}\rangle /dt=\text{Tr}(\dot{\rho}\hat{J}_{i})$ and the definition of $X=\langle \hat{J}_{x}\rangle
/j$, $Y=\langle \hat{J}_{y}\rangle /j$, and $Z=\langle
\hat{J}_{z}\rangle /j$ ($j=N/2$ is the scaling factor),
 we obtain the semiclassical equations of motion
\begin{eqnarray}
\dot{X} &=&2hY-\Gamma _{b}ZX-\gamma_\text{dep}X/2 \\
\dot{Y} &=&-2hX+2\lambda ZX-\Gamma _{b}ZY-\gamma_\text{dep}Y/2\\
\dot{Z} &=&-2\lambda XY+\Gamma _{b}(X^{2}+Y^{2}).
\end{eqnarray}
These three formulas   can not form a closed set of group,
and the constraint $X^{2}+Y^{2}+Z^{2}=1$ corresponds to the conservation of
angular momentum. Therefore, we can get a closed set of group with these four equations,
and obtain the steady-state analytical solutions or numerical solutions
from these equations. The numerical solutions for the finite $N$ will
lead to $\langle \hat{J}_{x}\rangle =\langle \hat{J}_{y}\rangle =0$
for all $\lambda $. Then we will derive the steady-state analytical solutions
of motion from these equations
\begin{eqnarray}
\label{ME6}2hY-\Gamma _{b}ZX-\gamma_\text{dep}X/2=0, \\
\label{ME7}-2hX+2\lambda ZX-\Gamma _{b}ZY-\gamma_\text{dep}Y/2=0, \\
\label{ME8}-2\lambda XY+\Gamma _{b}(X^{2}+Y^{2})=0, \\
\label{ME9}X^{2}+Y^{2}+Z^{2}=1.
\end{eqnarray}%
We solve the equations (\ref{ME6})-(\ref{ME9}) and find the critical point of the coupling strength $\lambda$,
\begin{equation}
\lambda _{c}=h+\frac{\Gamma _{b}^{2}}{4h}.
\end{equation}%
It is obviously that the critical point $\lambda _{c}$ is varying with $\Gamma_{b}$ and $h$, but immune to the dephasing $\gamma_\text{dep}$.

When the coupling strength satisfies $\lambda <\lambda _{c}$, the steady-state belongs
to the normal phase and the analytical solutions of motion are given by%
\begin{equation}\label{ME10}
Z_{np}=1,X_{np}=Y_{np}=0.
\end{equation}%
While for $\lambda >\lambda _{c}$, which is the region for the second-order phase transition, the
steady-state corresponds to the broken phase and the analytical solutions are given by%
\begin{eqnarray}
\label{ME11}Z_{bp} &=&Z_{0}-r_{0}, \\
\label{ME12}X_{bp} &=&\pm \sqrt{\dfrac{1-(Z_{0}-r_{0})^2}{1+\Gamma_{b}Z_{0}^2/2h}},\\
\label{ME13}Y_{bp} &=&\pm \dfrac{\Gamma _{b}}{2h}Z_{0}(\sqrt{\dfrac{1-(Z_{0}-r_{0})^2}{1+\Gamma_{b}Z_{0}^2/2h}}),
\end{eqnarray}%
where $Z_{0}=\dfrac{2h}{\Gamma_{b}^2}(\lambda -\sqrt{\lambda ^{2}-\Gamma _{b}^{2}(1+r_{0}\lambda/h )})$, and the dimensionless parameter $r_{0}=\gamma_\text{dep}/2\Gamma_{b}$.
The analytical solutions of motion exhibit a bifurcation at the critical
coupling strength $\lambda _{c}$.

In order to exhibit the macroscopic difference between the normal
phase and the broken phase, we can apply the Bloch vector to describe the
average collective spins. The NV center ensemble in this system can be interpreted as the
collective spin-$1/2$ particle system. According to the definition of
spin vacuum state and spin coherent state \cite{19}, the average value of the spin ensemble can be
expressed as $X=\sin \theta \cos \phi, Y=\sin \theta \sin \phi$, and $Z=\cos
\theta$, where the vector $(1,\theta, \phi )$ means the average-spin direction for this spin ensemble.
According to this definition, we can describe the macroscopic
eigenstate as $|\theta, \phi \rangle$. When we set $\theta =0$, the average value is $%
Z=1$, $X=0$, and $Y=0$ , which means the average-spin direction is along $+z$ axis and the spin ensemble belongs to the normal phase area.
On the other hand, if $\theta\neq 0$, the average value is $Z=\cos
\theta $, $X=\sin \theta \cos \phi$, and $Y=\sin \theta \sin \phi$, which means the Bloch
vector is rotated away from $+z$ axis and then the system falls in the broken phase area.
The second-order phase transition will occur in this area when $\lambda =\lambda _{c}$.
\begin{figure}
\centering \includegraphics[width=9cm]{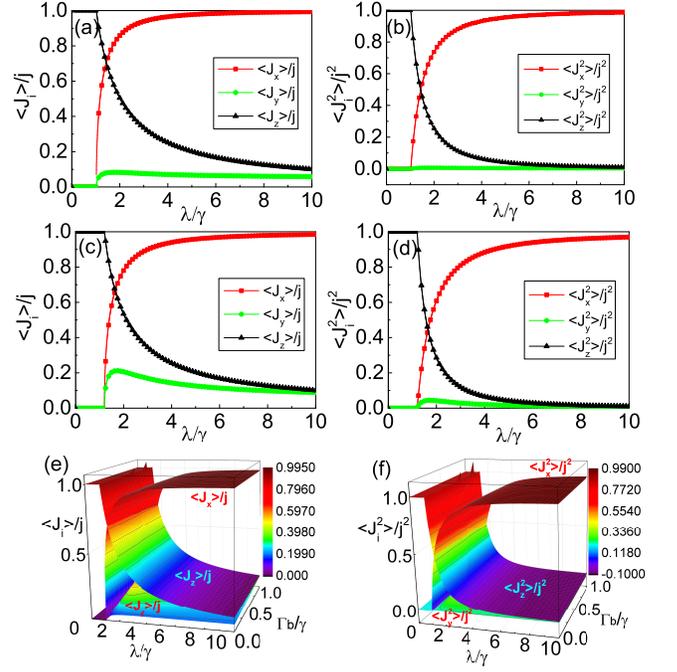}
\caption{\label{fig:wide}(Color online) The average values of the
first-order spin components and the quadratic spin components versus the dimensionless variables $\lambda/\gamma$ and $\Gamma _{b}/\gamma$,
where the parameters are chosen as $\gamma_\text{dep}=0.2\gamma$ and $h=\gamma$.
The average values of $\langle\hat{J}_{i}\rangle/j$ and $\langle\hat{J}_{i}^{2}\rangle/j^2$ versus $\lambda/\gamma$ with  $\Gamma _{b}=0.2\gamma$ for $(a)-(b)$  and $\Gamma_{b}=0.8\gamma$ for $(c)-(d)$.
$(e)-(f)$ The three dimensional plots of $\langle\hat{J}_{i}\rangle/j$ and $\langle\hat{J}_{i}^{2}\rangle/j^2$ versus ($\lambda/\gamma$, $\Gamma_{b}/\gamma$). }
\end{figure}

\begin{figure}
\centering \includegraphics[width=9cm]{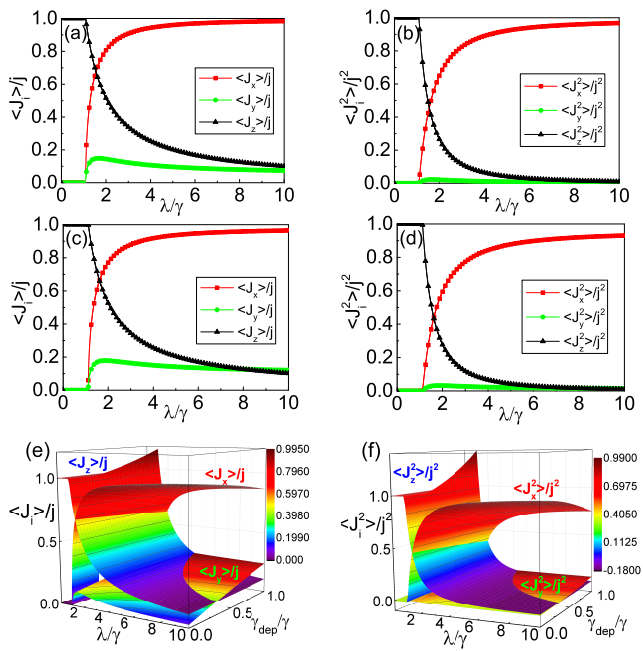}
\caption{\label{fig:wide}(Color online) The average values of the
first-order spin components and the quadratic spin components versus the dimensionless variables $\lambda/\gamma$ and $\gamma _\text{dep}/\gamma$,
where the parameters are chosen as $\Gamma_{b}=0.5\gamma$ and $h=\gamma$.
 The average values of $\langle\hat{J}_{i}\rangle/j$ and $\langle\hat{J}_{i}^{2}\rangle/j^2$ versus $\lambda/\gamma$ with $\gamma _\text{dep}=0.2\gamma$ for $(a)-(b)$ and $\gamma_\text{dep}=0.4\gamma$ for $(c)-(d)$.
$(e)-(f)$ The three dimensional plots of $\langle\hat{J}_{i}\rangle/j$ and $\langle\hat{J}_{i}^{2}\rangle/j^2$ versus ($\lambda/\gamma$, $\gamma_\text{dep}/\gamma$).}
\end{figure}

In this setup, we assume that the density of the NV centers is about $\varrho\sim 10^{15}\text{cm}^{-3}$ with the total number $N\sim 10^{12}$,
and we set the parameters as the last column for the simple LMG model in TABLE. \uppercase\expandafter{\romannumeral1}.
We can calculate the expectation values of the spin components of the Bloch vector numerically from the equations (\ref{ME6})-(\ref{ME9}),
where the effective coupling strengths $\lambda\sim 2\pi \times 0.25 $ MHz and $h\sim 2\pi \times 0.25 $ MHz, the effective dissipation rate  $\Gamma_{b}\sim 2\pi \times 0.05 $ MHz and the dephasing rate $\gamma_\text{dep}\sim 2\pi \times 0.02$ MHz. The scaling  factor is $\gamma\sim 2\pi \times 0.25$ MHz.

In order to illustrate the effects  of the coupling $\lambda$, the dissipation $\Gamma_{b}$ and the dephasing $\gamma_\text{dep}$ on   phase transitions,  in Figs.~2 and  3  we present the average values of the first-order
spin components and the quadratic spin components under different conditions from equations (\ref{ME10})-(\ref{ME13}).

As shown in Fig.~2(a), (b), (c) and (d), we set $\gamma_\text{dep}=0.2\gamma$ and $h=\gamma$, and get the curve for the average values of the first-order $\langle\hat{J}_{i}\rangle/j$ and the quadratic $\langle\hat{J}_{i}^{2}\rangle/j^2$ spin components varying with $\lambda/\gamma$,
where $\Gamma_{b}=0.2\gamma$ (Fig. 2 (a) and (b)) and $\Gamma_{b}=0.8\gamma$ (Fig.~2(c) and (d)).
In Fig.~2(e) and (f), we also display the three dimensional surface of these average values varying with the two parameters ($\lambda/\gamma$, $\Gamma_{b}/\gamma$) when $\gamma_\text{dep}=0.2\gamma$ and $h=\gamma$.
One can find that the second-order phase transition occurs at the point near $\lambda _{c}$.
It is shown that the average value of $\langle\hat{J}_{z}\rangle/j$ or $\langle\hat{J}_{z}^{2}\rangle/j^2$ is always $1$ when $\lambda <\lambda _{c}$, and then displays a discontinuous transition as $\lambda /\gamma$ increases to a value larger than 1. As for the $\langle\hat{J}_{x}\rangle/j$ or $\langle\hat{J}_{x}^{2}\rangle/j^2$ component,
it exhibits a reversed behavior compared to the $\langle\hat{J}_{z}\rangle/j$ or $\langle\hat{J}_{z}^{2}\rangle/j^2$ component.
However, both display the discontinuous behaviour at the point $\lambda_{c}$, which signifies quantum phase transition occurs.
In addition, as shown in Fig.~2(e) and (f), when the dissipation $\Gamma_{b}$ increases,
 the critical point $\lambda_{c}$ will be slightly shifted.
The value of $\langle\hat{J}_{y}\rangle/j$ or $\langle\hat{J}_{y}^{2}\rangle/j^2$  in the vicinity of $\lambda_{c}$ also increases a little.
Comparing Fig.~2(a) or (b) with Fig.~2(c) or (d),
the critical point has changed from about $\lambda_{c}=1.01\gamma$ to $\lambda_{c}=1.16\gamma$.
Moreover, in the vicinity of $\lambda_{c}$,
the value of $\langle\hat{J}_{y}\rangle/j$ has changed from a value  less than $0.1$ to one larger than $0.2$.

In Fig.~3(a), (b), (c) and (d), we set $\Gamma_{b}=0.5\gamma$ and $h=\gamma$,
and can obtain the illustrations for the values of the first-order $\langle\hat{J}_{i}\rangle/j$ and the quadratic $\langle\hat{J}_{i}^{2}\rangle/j^2$ spin components varying with $\lambda/\gamma$,
where $\gamma_\text{dep}=0.2\gamma$ (Fig.~3a) and (b)) or $\gamma_\text{dep}=0.4\gamma$ (Fig.~3(c) and (d)).
Moreover, as shown in Fig.~3(e) and (f), we also plot the three dimensional surface of these average values varying with  the parameters
($\lambda/\gamma$, $\gamma_{dep}/\gamma$) when $\Gamma_{b}=0.5\gamma$ and $h=\gamma$.
We can obtain the  critical point for the phase transition $\lambda_{c}=1.0625\gamma$ from Fig.~3,  which is immune to the dephasing $\gamma_\text{dep}$.
Figure.~3 displays the same behavior of the average spin components as in Fig. ~2.
We find a discontinuous behavior as $\lambda$ increases to a value larger than $\lambda_{c}$,
and $\langle\hat{J}_{x}\rangle/j$ or $\langle\hat{J}_{x}^{2}\rangle/j^2$
exhibits a reversed behavior compared to $\langle\hat{J}_{z}\rangle/j$ or $\langle\hat{J}_{x}^{2}\rangle/j^2$.
Although the dephasing will not affect the critical point, we can not neglect its effect on the phase transition.
As shown in Fig.~3(e) and (f), when increasing $\lambda$ and $\gamma_\text{dep}$,
the values of $\langle\hat{J}_{y}\rangle/j$ will be enlarged continuously.
Meanwhile $\langle\hat{J}_{x}\rangle/j$ will be suppressed continuously too.
When $\lambda/\gamma\sim10$ and $\gamma_\text{dep}/\gamma\sim1$,
$\langle\hat{J}_{y}\rangle/j$ will keep  the value of $ 0.24$,
while $\langle\hat{J}_{x}\rangle/j$ will be about $ 0.87$.
Therefore,  we can simulate the
second-order quantum phase transition with this hybrid quantum system, which provides the very convincing evidence for manipulating the spin ensembles  realistically in our scheme.

\section{The spin squeezed state}

It is known that spin squeezed states can be prepared efficiently by one-axis twisting or two-axis
counter-twisting interactions \cite{19}. These kinds of interactions are equivalent to the LMG model
with the form of equation (\ref{ME4}) or (\ref{ME3}). Although the two-axis
counter-twisting Hamiltonian is superior to the one-axis twisting one, the spin-spin interaction with the form of two-axis counter-twisting has not been realized in any experiments due to the demanding requirements \cite{19,20,21,22,61}.
We will discuss how to prepare spin squeezed states by simulating the two-axis
counter-twisting interaction with equation (\ref{ME3}) in our scheme.
In this system, the coherent coupling strength between a single NV center and the cavity is much less than
the cavity dissipation rates $(\vert g_{i}\vert \ll \kappa _{a,b})$, but this collective coupling can be enhanced by increasing the number of the NV centers. Therefore,
we can neglect the dissipations as long as the condition $\sqrt{N}\vert g_{i}\vert\gg\kappa _{a,b}$ is satisfied.

In order to describe the degree of spin squeezing, we introduce the definition \cite{61}
\begin{equation}
\xi ^{2}=\frac{4\min (\Delta \hat{J}_{\vec{n}\perp }^{2})}{N},
\end{equation}%
where $\vec{n}_{\perp }$ refers to an axis perpendicular to the
mean-spin direction, and the term \textquotedblleft min\textquotedblright\ is the minimization over all directions
$\vec{n}_{\perp }$. The first step is to determine the mean-spin direction $\vec{n}_{0}$ by the
expectation values $\langle \hat{J}_{\alpha }\rangle$, with $\alpha \in
\{x,y,z\}$.  We write $\vec{n}_{0}$ with spherical coordinates $%
\vec{n}_{0}=(\sin \theta \cos \phi, \sin \theta \sin \phi, \cos \theta )$,
and this description is equivalent to the coherent spin state $|\theta,\phi \rangle$.
We can get the other two orthogonal bases which are perpendicular to $\vec{n}_{0}$,
\begin{eqnarray}
\vec{n}_{1} &=&(-\sin \phi ,\cos \phi ,0), \\
\vec{n}_{2} &=&(\cos \theta \cos \phi ,\cos \theta \sin \phi ,-\sin \theta ).
\end{eqnarray}%
Hence, $\vec{n}_{\perp }=\vec{n}_{1}\cos \beta +\vec{n}_{2}\sin \beta$ is the arbitrary direction vector perpendicular to $\vec{n}_{0}$, and we
can find a pair of optimal quadrature operators by tuning $\beta$. Then we obtain two components of the angular momentum,
\begin{eqnarray}
\hat{J}_{\vec{n}_{1}} &=&-\sin \phi \hat{J}_{x}+\cos \phi \hat{J}_{y}, \\
\hat{J}_{\vec{n}_{2}} &=&\cos \theta \cos \phi \hat{J}_{x}+\cos \theta \sin \phi
\hat{J}_{y}-\sin \theta \hat{J}_{z}.
\end{eqnarray}%
As a result, we acquire the expression of the optimal squeezing parameter
\begin{equation}
\xi ^{2}=\frac{2}{N}[\langle \hat{J}_{\vec{n}_{1}}^{2}+\hat{J}_{\vec{n}%
_{2}}^{2}\rangle -\sqrt{(\langle \hat{J}_{\vec{n}_{1}}^{2}-\hat{J}_{\vec{n}%
_{2}}^{2}\rangle )^{2}+4\text{Cov}(\hat{J}_{\vec{n}_{1}},\hat{J}_{\vec{n}_{2}})}],
\end{equation}%
where
\begin{equation}
\text{Cov}(\hat{J}_{\vec{n}_{1}},\hat{J}_{\vec{n}_{2}})=\frac{1}{2}\langle \hat{J}_{\vec{n}_{1}}\hat{J}_{\vec{n}_{2}}+\hat{J}_{\vec{n}_{2}}\hat{J}_{\vec{n}_{1}}\rangle \notag.
\end{equation}
We can distinguish between spin coherent states and spin squeezed states distinctly for this NV center ensemble according to $\xi ^{2}=1$ or $\xi ^{2}<1$.
Therefore, it is imperative for us to carry out some numerical calculations for the squeezing parameter in this system.

It is evident that one can strengthen the spin squeezing degree by increasing the total number of the spins \cite{20,21,22,45,61}.
As discussed  above, we can get the relations $\sigma _{a}\alpha _{a}\simeq \sigma _{a}\beta _{a}$,
$\sigma _{b}\alpha _{b}\simeq -\sigma _{b}\beta _{b}$ by tuning the collective couplings, the Rabi frequencies, and the detunings.
Then we can get $\alpha=\beta=\sqrt{2}/2$ when setting $\sigma _{a}\simeq \sqrt{2}\sigma _{a}\alpha _{a}$, $\sigma _{b}\simeq \sqrt{2}\sigma _{b}\alpha _{b}$. We investigate equations (\ref{ME3}) and (\ref{ME15}) and assume that the coefficients are $\lambda\equiv|\Lambda_{a}|=|\Lambda_{b}|$, $\sigma _{a}\simeq\sigma _{b}\equiv\sigma$, $\zeta _{a}\simeq \zeta _{b}\equiv\zeta$, and the dissipations of the supermodes are $\kappa _{a}=\kappa _{b}\equiv\kappa$.

We choose the first column parameters for the conventional LMG model in TABLE. \uppercase\expandafter{\romannumeral1}
and assume the dissipations $\kappa\sim 2\pi\times 0.1$ MHz.
We choose the number of NV centers $N \simeq 10^{6}$ and get the collective coupling $\sqrt{N}g_{i}\simeq 2\pi\times12$ kHz.
Then the effective coupling coefficient is $\lambda=\sigma^{2}|\zeta|/(\kappa^{2}+\zeta^{2})\simeq\sigma^{2}|\zeta|/\kappa^{2}\simeq 0 $ Hz
and the effective dissipations are $\Gamma _{a,b}=\sigma^{2}\kappa /(\kappa^{2}+\zeta^{2})\simeq \sigma^{2}/\kappa\simeq 2\pi\times 1.8$ Hz.
So we can not prepare the spin squeezed state when $N\simeq 10^{6}$,  because  $\lambda$ and $ \Gamma _{a,b} $ are too weak.
When we set $N \simeq 10^{10}$ and the collective coupling as $\sqrt{N}g_{i}\simeq 2\pi\times 1.2$ MHz,
utilizing the same parameters above, we can get the effective coupling coefficient as $\lambda=\sigma^{2}|\zeta|/(\kappa^{2}+\zeta^{2})\simeq 2\pi\times 4.43 $ kHz
and the effective dissipations as $\Gamma _{a,b}=\sigma^{2}\kappa /(\kappa^{2}+\zeta^{2})\simeq 2\pi\times 16.4 $ kHz.
It is evident that we can not prepare the spin squeezed state unless $N > 10^{10}$.
On the other hand, if we choose $N \simeq 10^{12}$ and utilize the same parameters above, we find that,
the effective coupling $\lambda\sim 2\pi\times 65.2$ kHz is much stronger than the effective dissipations $\Gamma_{a,b}\sim 2\pi\times 2.4$ kHz.
Therefore, one can not prepare the spin squeezed state unless $N>10^{10}$ in this setup.
These estimations above also show that the appropriate choice for simulating spin squeezed state is $N\simeq 10^{12}$.

In the condition of weak excitations and $N\gg 1$, we map the collective spin operators
$\hat{J}_{+}(\hat{J}_{-})$ into the boson operators $\hat{d}^{\dagger }(\hat{d})$ in the
Holstein-Primakoff representation,
\begin{eqnarray}
\hat{J}_{+} &=&\sqrt{N}\hat{d}^{\dagger },  \notag \\
\hat{J}_{-} &=&\sqrt{N}\hat{d}, \\
\hat{J}_{z} &=&(\hat{d}^{\dagger }\hat{d}-\frac{N}{2}),  \notag
\end{eqnarray}%
where the operators $\hat{d}$ and $\hat{d}^{\dagger }$ obey the standard boson
commutator $[ \hat{d},\hat{d}^{\dagger }] =1$.
The equations (\ref{ME3}) and (\ref{ME15}) can be transformed as
\begin{equation}\label{ME14}
\dot{\rho}=-i[H_{T},\rho ]+\Gamma_{a}D[\hat{d}]\rho+\Gamma_{b}D[\hat{d}^{\dag }]\rho+\gamma_\text{dep}D[\hat{d}^{\dag }\hat{d}]\rho,
\end{equation}%
\begin{equation}
H_{T}=-2h\hat{d}^{\dag }\hat{d}-\lambda \hat{d}^{2}-\lambda \hat{d}^{\dag 2}.
\end{equation}%

\begin{figure}
\centering \includegraphics[width=9cm]{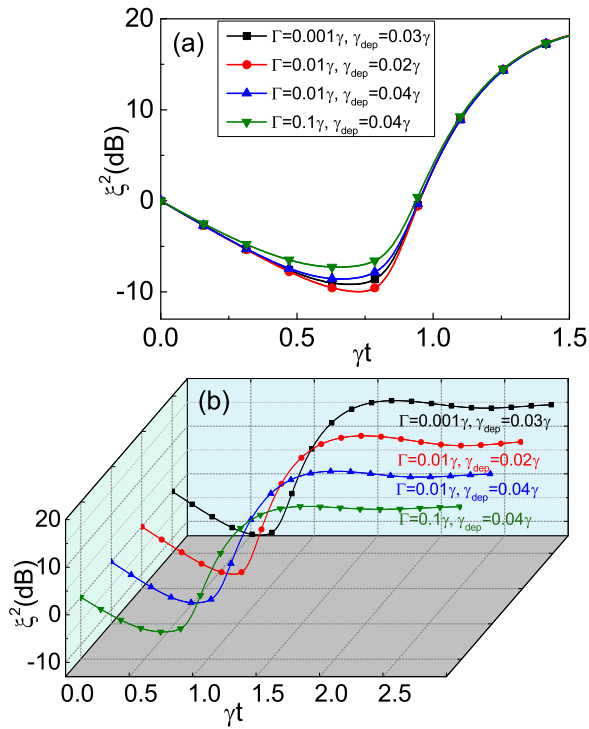}
\caption{\label{fig:wide}(Color online) Time evolution of $\xi ^{2}$ with $H_{T}$ for $N\simeq 10^{12}$,
where the parameters are chosen as $\lambda/\gamma\simeq1$, $2h/\gamma\simeq 0$
and the different effective dissipations are $\Gamma_{a}=\Gamma_{b}=\Gamma=0.1\gamma, 0.01\gamma, 0.001\gamma$ and the dephasing rates are $\gamma_{dep}=0.02\gamma, 0.03\gamma, 0.04\gamma$ respectively. $(a)$ The evolution for a shorter time $0 \leq t \leq 1.5/\gamma$.
$(b)$ The evolution for a longer time $0 \leq t \leq 3/\gamma$.}
\label{fig.level}
\end{figure}
We assume the collective spins are initially prepared in the state $|\theta=0 \rangle$ ($Z=1$) and the dissipations satisfy $\Gamma_{a}=\Gamma_{b}$.
Then we solve the master equation (\ref{ME14}) numerically and present the simulations in Fig. 4.
In this numerical simulation, the parameters are  set as those in the first column of TABLE. \uppercase\expandafter{\romannumeral1}. With these parameters, we can obtain the effective coupling coefficient $\lambda \simeq 2\pi \times 65 $ kHz,
and the corresponding effective dissipations $\Gamma_{a,b} \simeq 2\pi \times 0.065 $ kHz, $ 2\pi \times 0.65 $ kHz and $2\pi \times 6.5 $ kHz.
Here the dephasing rates are $\gamma_\text{dep}\simeq 2\pi \times 1.3$ kHz, $2\pi \times 1.95$ kHz, and $2\pi \times 2.6$ kHz.
Therefore, the scaling factor in Fig. 4 is $\gamma \simeq 2\pi \times 65 $ kHz.

As shown in Fig.~4(a), we can definitely get spin squeezed  for a relatively short evolution time.
The system can always be in the spin squeezed state when $0<t\leq\frac{0.8}{\gamma}$,
and the squeezing parameter is about $\xi ^{2}\sim -10$ dB when $\Gamma_{a,b}\simeq 2\pi\times 0.065 $ kHz and $\gamma_\text{dep}\simeq 2\pi \times 1.3$ kHz.
Nevertheless, this nonclassical state will be destroyed  because of the dissipation and dephasing.
In Fig.~4(b), it is evident that the dissipation and dephasing will reduce the squeezing degree,
and it finally evolves into a  thermal state.

To examine the feasibility of our scheme in realistic experiment, we now
discuss the relevant available experimental parameters. The magnetic
coupling strength between the cavity and a single NV center is about $%
g_{i}\sim 2\pi\times 10$ Hz. The dissipation rate of microwave cavities is about $\kappa
_{a},\kappa _{b}> 2\pi\times 1$ kHz. In the practical situation, we consider a
spin ensemble of $N\sim10^{12}$ NV centers coupled to the
cavity, and the collective coupling strength satisfies $\sqrt{N}g_{i}\sim
2\pi\times 10$ MHz \cite{52}. The Rabi frequency is about $|\Omega^{*} _{i}|\sim 2\pi\times 1$ MHz.
Based on the chosen parameters,  the time for maintaining the spin squeezed state
is about $0 < t \leq 2 $ $\mu s$ in this system, and we can obtain the degree of
squeezing $\xi ^{2}\sim-10$ dB with the relative smaller number of the
excitations. On the other hand, the relaxation time of the NV spin triplet
ranging from milliseconds at room temperature to several seconds at
low temperature has been reported \cite{53}.
In addition, the dephasing time $(T_{2}\propto 1/\gamma_\text{dep})$
more than $400$ $\mu $s for spin ensemble has been demonstrated,
and can be raised to about $2$ ms with an isotopically pure diamond sample \cite{62,63,64,65,66}.
Thus, the coherence time is sufficient for achieving the desired spin squeezed state.

\section{Conclusion}

In summary, with the assistance of classical microwave fields and superconducting coplanar waveguide supermodes,
we implement an exquisite setup for simulating LMG model with an NV center ensemble in diamond.
Utilizing this hybrid quantum system,
we can not only achieve the second order quantum phase transition when the
coefficient satisfies $\lambda =\lambda _{c}$,
but also prepare the spin squeezed state via the LMG Hamiltonian with the form of two-axis counter-twisting spin-spin interactions.
In this protocol, due to the weak single coupling coefficient $g_{i}\ll\kappa_{i}$,
the two-axis counter-twisting interactions will not be valid unless $N> 10^{10}$.
Moreover, in spite of the  negative effect of dissipations and dephasing,
the squeezing degree is near to $-10$ dB when we choose $N\simeq 10^{12}$.
We can further enhance the squeezing degree by increasing the number of the NV centers.
Furthermore, if we extend this scheme to the LMG \textquotedblleft  Lattice \textquotedblright\ model,
there will be  more  physics. This scheme is a new attempt for utilizing the hybrid quantum system to simulate the LMG model.
\section*{Acknowledgments}

This work is supported by the NSFC under
Grant Nos. 11774285, 11474227 and 11534008,  as well as  the Fundamental Research Funds for the
Central Universities. Part of the simulations are coded
in PYTHON using the QUTIP library \cite{67,68}.

\end{document}